\documentclass[aps,prl,twocolumn,superscriptaddress,noshowpacs]{revtex4}
\usepackage{bm}
\usepackage[utf8]{inputenc}
\usepackage{physics}
\usepackage[english]{babel} 
\usepackage{amsmath,amsfonts,amssymb}
\usepackage{graphicx}
\usepackage{color}

\begin{document}
\title{Universal semiclassical equations based on the quantum metric}
\author{C.~Leblanc}
\affiliation{Institut Pascal, PHOTON-N2, Universit\'e Clermont Auvergne, CNRS, SIGMA Clermont, F-63000 Clermont-Ferrand, France.}

\author{G.~Malpuech}
\affiliation{Institut Pascal, PHOTON-N2, Universit\'e Clermont Auvergne, CNRS, SIGMA Clermont, F-63000 Clermont-Ferrand, France.}

\author{D.~D.~Solnyshkov}
\affiliation{Institut Pascal, PHOTON-N2, Universit\'e Clermont Auvergne, CNRS, SIGMA Clermont, F-63000 Clermont-Ferrand, France.}
\affiliation{Institut Universitaire de France (IUF), F-75231 Paris, France}

\begin{abstract}
We derive semiclassical equations of motion for an accelerated wavepacket in a two-band system. We show that these equations can be formulated in terms of the static band geometry described by the quantum metric. We consider the specific cases of the Rashba Hamiltonian with and without a Zeeman term. The semiclassical trajectories are in full agreement with the ones found by solving the Schr\"odinger equation. This formalism successfully describes the adiabatic limit and the anomalous Hall effect traditionally attributed to Berry curvature. It also describes the opposite limit of coherent band superposition giving rise to a spatially oscillating Zitterbewegung motion. At $k=0$, such wavepacket exhibits a   circular  trajectory  in  real  space,  with its radius given by the square root of the quantum metric. This quantity appears as a universal length scale, providing a geometrical origin of the Compton wavelength.
\end{abstract}

\maketitle
General relativity is the first example of a geometrical theory of motion, where the particle trajectories are not governed by gravitational forces, but are found as the geodesics of the spacetime metric.  In a completely different perspective, the semiclassical theory of electron dynamics in solids was derived in the 1930s from quantum mechanics \cite{Jones1934,Zak1968}, involving as a key element the wavevector-dependent group velocity. These equations have been corrected in 1999 \cite{Sundaram1999} to include the impact of the Berry curvature and describe the anomalous Hall effect (AHE). The AHE was discovered in the 50s \cite{Luttinger1954}, but the deep understanding of the underlying physics and of its importance came with its description in terms of geometrical properties of the quantum space.

Indeed, the geometry of the quantum space is actively studied since the 1980s \cite{provost1980riemannian,berry1984quantal,berry1989quantum}, bringing the  description of the quantum Hall effect \cite{Klitzing1980,Thouless1982}, and the forthcoming enormous development of topological physics \cite{Hasan2010,lu2014topological,ma2019topological,solnyshkov2021microcavity}. The geometrical information about the eigenstates of a Hamiltonian is contained in the gauge-invariant quantum geometric tensor, whose symmetric real part  defines the quantum metric (QM) characterizing distances between states~\cite{provost1980riemannian} in a parameter space. Its antisymmetric imaginary part determines the Berry curvature~\cite{berry1984quantal}.
The key hypothesis of the Niu-Sundaram \cite{Sundaram1999,Chang2008} equations is the adiabatic approximation, when the wavepacket remains in a single energy band, as in the original work of Berry \cite{berry1984quantal}.
The extension to the situation where several bands have comparable populations was done in Ref.~\cite{Culcer2005}, but using time-dependent components of the generalized Berry curvature tensor, which depend on initial conditions and not only on static band parameters.

The potential roles of the QM have been more recently underlined in the calculations in  quantum informatics, quantum phase transitions \cite{Zanardi2007}, magnetic susceptibility \cite{Gao2014,Piechon2016}, excitonic levels \cite{PhysRevLett.115.166802}, superfluidity in flat bands \cite{peotta2015superfluidity,Liang2017}. The QM is now explicitly accounted for in the design and engineering of topological systems \cite{Kremer2019,Salerno2020}, and its integral is linked with the Chern number \cite{Roy2014,jackson2015geometric,Piechon2016,mera2021k}. Experimental measurements of the QM in different systems also start to appear \cite{Tan2019,Yu2019,gianfrate2020measurement}.
In particular, it is well understood that the QM should appear in the description of transitions between quantum levels. For example, it allows to describe small non-adiabatic corrections to the AHE \cite{Gao2014,Bleu2018,Holder2020}.
In some cases, the QM was even found to dominate the dynamics. This occurs in systems with non-reciprocal directional dichroism \cite{Gao2019} and also in strongly non-Hermitian systems in vicinity of the exceptional points, where the evolution can never be adiabatic \cite{NH1}.
A situation of a particular interest occurs in spin-orbit-coupled systems \cite{Shen2005,Jin2006,Hatano2007,Yang2008} which can be described in terms of  non-Abelian gauge potentials \cite{Yang1954,Ryder} with emergent vectorial charges \cite{Tercas2014,Chen2019}. The resulting Zitterbewegung (ZBW)  motion \cite{Yang2019,fieramosca2019chromodynamics} involves a coherent superposition of several bands. The ZBW is studied theoretically and experimentally in various electronic \cite{Bolte2007,Zawadzki2011,Tarasenko2018}, atomic \cite{gerritsma2010quantum,LeBlanc2013,Qu2013}, and photonic systems \cite{Dreisow2010,Guo2015,Guo2016,Silva2019} including polaritons  \cite{Sedov2018, Sedov2020}. This is an appealing situation for its description in terms of QM, as noticed in \cite{Iskin2019}, where the QM was shown to be responsible for a contribution to the effective mass.

In this work, we derive semiclassical equations of motion in a two-band system using only the static band geometry encoded in the QM. The solutions of these  new equations are in complete agreement with the direct numerical solutions of the Schr\"odinger equations for all the cases we considered. They describe the AHE, traditionally attributed to Berry curvature. They also describe the opposite limit, when the wavepacket is coherently distributed over the two bands, and in particular the ZBW motion induced by an emergent non-Abelian gauge field.  We show that a wavepacket centered at $k=0$ exhibits a circular  trajectory  in  real  space,  with its radius given by the square root of the quantum metric. This quantity appears as a universal length scale, determining the uncertainty of the position of a particle involving several bands. It provides a geometrical origin of the Compton wavelength.

\begin{figure}[tbp]
\begin{center}
\includegraphics[width=0.95\linewidth]{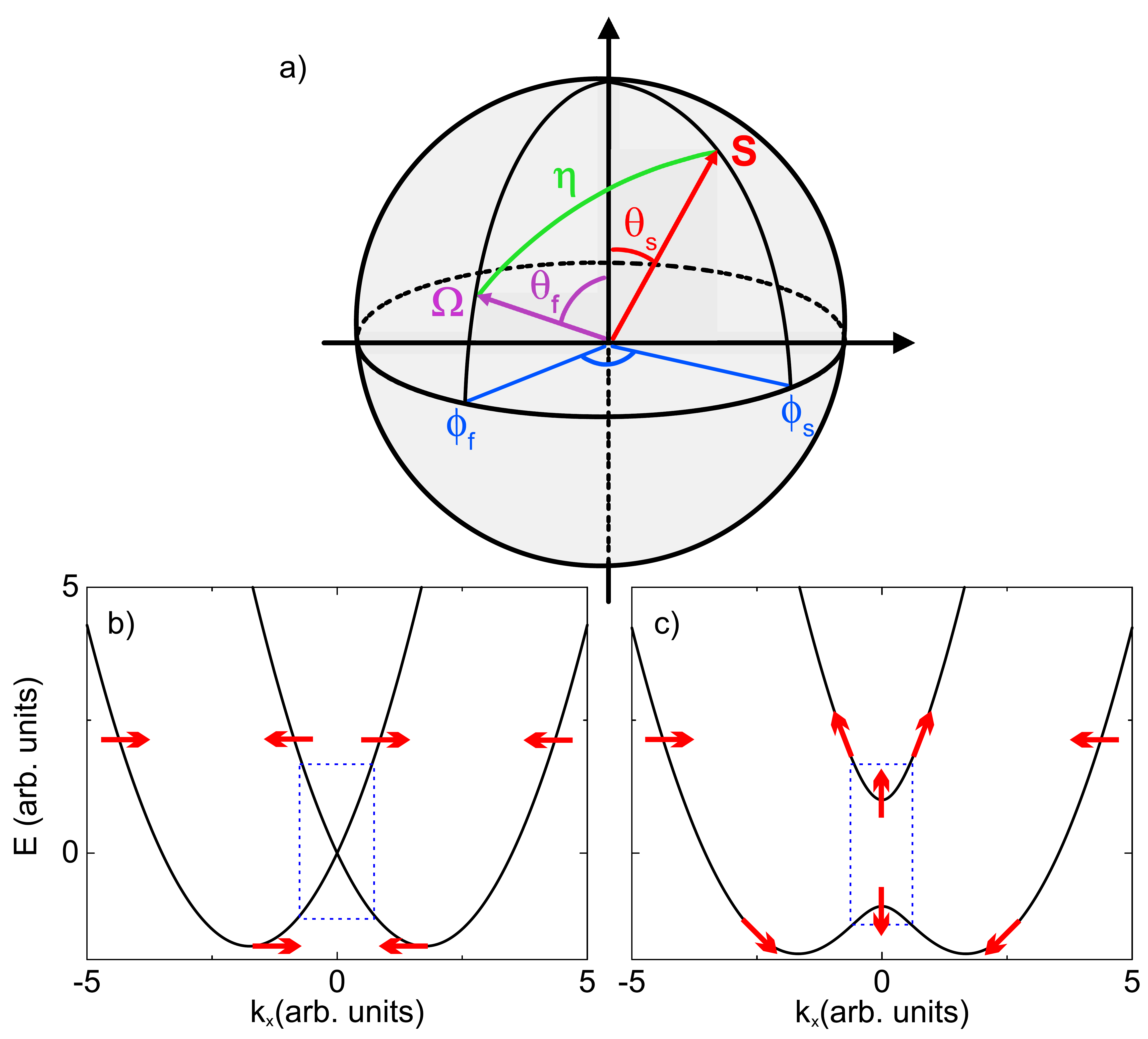}
\caption{a) Bloch sphere representation showing the pseudospin $\mathbf{S}$, the effective field $\mathbf{\Omega}$, and their polar and azimuthal angles $\theta$ and $\phi$. The angle between the spin and the field is $\eta$. b,c) Dispersion along $k_x$ of the eigenmodes of the b) Rashba  and c) Rashba+Zeeman  Hamiltonians. The red arrows show the pseudo-spin orientation of the modes. \label{fig1}}
\end{center}
\end{figure}

\emph{The model} We begin with the Hamilton's equations of motion for a wavepacket. Working with a 2-band system allows us to use the mapping to the pseudospin $\bm{S}$ interacting with effective fields \cite{Feynman1957}, and the associated geometry of the Bloch sphere (Fig.~\ref{fig1}). A general superposition of two eigenstates can be written as
\begin{equation}
\left| \psi  \right\rangle  = {c_1}\left| {{\psi _1}} \right\rangle  + {c_2}\left| {{\psi _2}} \right\rangle  = \left| {\begin{array}{*{20}{c}}
{{c_1}}\\
{{c_2}}
\end{array}} \right\rangle  = \left| {\begin{array}{*{20}{c}}
{\cos \frac{\theta_s }{2}{e^{ - i\phi_s }}}\\
{\sin \frac{\theta_s }{2}}
\end{array}} \right\rangle,
\end{equation}
where $\theta_s$ and $\phi_s$ are the time-dependent angles, giving the orientation of the pseudospin on the Bloch sphere. The equations of motion for the spatial degrees of freedom are therefore accompanied with the precession equation for the pseudospin describing the wavepacket distribution within the two bands:
\begin{equation}
    \Dot{\bm{p}}=-\frac{\partial H}{\partial \bm{r}},\quad \Dot{\bm{r}}=\frac{\partial H}{\partial \bm{p}}, \quad \Dot{\bm{S}}=\mathbf{S}\times\bm{\Omega}\label{Hameqns}
\end{equation}
Here, $\bm{r}$ is the spatial coordinate of the wavepacket center of mass, $\bm{p}=\hbar\bm{k}$ is the center of mass momentum of the wavepacket ($\bm{k}$ is the center of mass wavevector). At a given moment of time, the effective field is $\mathbf{\Omega}(\mathbf{k}(t))$ and the pseudospin is $\mathbf{S}(t)$, shown in Fig.~\ref{fig1} with violet and red arrows, respectively.  While it is often possible to convert Hamilton's equations to geodesics equation in an abstract metric \cite{casetti2000geometric}, our goal is rather to elucidate the role of the QM, while keeping the other coordinates intact.

The first of Eqns.~\eqref{Hameqns} describes the acceleration of the wavepacket due to a spatial gradient of the potential. We will rather focus on the second of Eqns.~\eqref{Hameqns}, describing the group velocity. The Hamilton's function $H$ corresponds to the energy $E=\bra{\psi}\hat{H}\ket{\psi}$ of the full wavepacket. It depends on the wavevector both directly, via the band dispersion $E_i(\mathbf{k})$, and indirectly, via the fractions $f_i=|c_i|^2$. This energy can be rewritten as $    E=f_1 E_1+f_2 E_2=(f_1+f_2)\Bar{E}+(f_2-f_1)\hbar\Omega$
where $\Bar{E}$ is the spinless part of the dispersion and $\Omega$ is the absolute value of the effective field.

We characterize the pseudospin $\mathbf{S}$ and the effective field $\mathbf{\Omega}$ by the respective spherical coordinates $\theta_s,\phi_s$ and $\theta_f,\phi_f$ (see Fig.~\ref{fig1}). The fractions $f_i$ are determined by the distance between these two vectors on the Bloch sphere $\eta$ as $f_{1,2}=\left(1\pm\cos\eta\right)/2$,
and this distance can be found from the spherical law of cosines \cite{suppl}. Our key idea is to use the QM as the link between the angles on the Bloch sphere and the wavevectors. By definition, the QM provides a link between the quantum distance $ds$ and the distance in reciprocal space:
\begin{equation}
 ds^2=g_{k_i,k_j} dk_i dk_j,
\end{equation}
with the QM defined by
\begin{equation}
    g_{ij}=\Re\left[\bra{\frac{\partial\psi_1}{\partial k_i}}\ket{\frac{\partial \psi_1 }{\partial k_j}}-\bra{\psi_1}\ket{\frac{\partial \psi_1 }{\partial k_i}}\bra{\frac{\partial\psi_1 }{\partial k_j}}\ket{\psi_1}\right]
    \label{gij}
\end{equation}    
The corresponding quantum distance for the displacement $d\eta$ on the  Bloch sphere is $ds^2=(d\eta)^2/4$. Ultimately, the equations of motion read (see \cite{suppl} for details):
\begin{equation}
    \hbar\mathbf{\Dot{k}}=-\frac{\partial E}{\partial \mathbf{r}},\quad \Dot{\mathbf{S}}=\mathbf{S}\times\mathbf{\Omega}\label{dsdqdt}
\end{equation}
\begin{widetext}
\begin{eqnarray}
    \hbar \mathbf{\Dot{r}}&=&\frac{\partial \Bar{E}}{\partial \mathbf{k}}-2\frac{\partial\hbar\Omega}{\partial \bm{k}}\left(\cos\theta_s\cos\theta_f+\sin\theta_s\sin\theta_f\cos\left(\phi_f-\phi_s\right)\right)\nonumber\\
    &-&\hbar\Omega\sqrt{g_{\mathbf{kk}}}\frac{\left[\left(-\cos\theta_s\sin\theta_f+\sin\theta_s\cos\theta_f\cos(\phi_f-\phi_s)  \right)- \sin\theta_s\sin(\phi_f-\phi_s)\sin\theta_f\left(\frac{\partial \phi_f}{\partial \theta_f}\right) \right]}{\sqrt{1+\sin^2\theta_f\left(\frac{\partial \phi_f}{\partial \theta_f}\right)^2}}\label{longeq}
\end{eqnarray}
\end{widetext}

In these expressions, the QM $\sqrt{g_{\bm{kk}}}=(\sqrt{g_{k_x,k_x}},\sqrt{g_{k_y,k_y}})^T$ is that of a single band (the lowest energy band of the doublet). We see that the QM appears as a overall factor of the corresponding term, entering~\eqref{longeq} together with $\Omega$ and thus completely determining the scale of the corresponding physical effect. The physical meaning of this term is the modification of the energy of the wavepacket due to its redistribution over the two bands with the rotation of the spin, and this is controlled by the QM.

Although this equation does not include the Berry curvature explicitly, it allows to recover the semiclassical equations of Ref.~\cite{Sundaram1999} with the Berry curvature terms in the adiabatic limit (see \cite{suppl}). In spite of being written only in terms of the QM, it entirely describes the AHE drift, and allows to go far beyond it, as we shall show below. Another advantage of Eq.~\eqref{longeq} is that it contains only the static properties of the bands.

If the Hamiltonian is such that the effective field remains for all $k$ in the equatorial plane ($\theta_f=\pi/2$, $\partial\phi_f/\partial\theta_f=\infty$), as is the case for the massless Dirac, Rashba, Dresselhaus, and TE-TM \cite{Kavokin2005} Hamiltonians, Eq.~\ref{longeq} is considerably simplified, reducing to
\begin{equation}
    \hbar \mathbf{\Dot{r}}=\frac{\partial E}{\partial \mathbf{k}}+\hbar\Omega\sqrt{g_{\bm{kk}}}
    \sin\theta_s\sin(\phi_f-\phi_s)\label{longeq2}
\end{equation}
with $E=\Bar{E}+\hbar\Omega(\cos\theta_s\cos\theta_f+\sin\theta_s\sin\theta_f\cos\left(\phi_f-\phi_s\right))$.
In what follows, we consider a Rashba Hamiltonian extensively studied in electronics, spintronics, and photonics, both with and without a Zeeman field: 
\begin{equation}
 \hat{H} = \frac{{{\hbar ^2}{k^2}}}{{2m}} + \alpha \bm{k}\cdot\bm{\sigma} +  \Delta{\sigma _z}
\label{RashbaHam}
\end{equation}
where $\bm{\sigma}$ is a vector of Pauli matrices,  $\alpha$ is the Rashba  magnitude, and $\Delta$ the magnitude of the effective Zeeman field. When $\Delta=0$, the eigenvalues are ${E_{\pm }} = \hbar ^2 k^2/(2m) \pm \alpha  k$, plotted in Fig 1(b). Close to $k=0$, the Hamiltonian is analogous with a 2D massless Dirac Hamiltonian, with $\alpha$ playing the role of the speed of light $c$. The bands have no distributed Berry curvature. A non-zero Zeeman field opens a gap at $k=0$ [Fig 1(c)], making appear an effective mass $m_{eff}=\hbar^2\Delta^2/\alpha^2$ (equivalence with a massive Dirac Hamiltonian). The corresponding bands show a non-zero distributed Berry curvature (see \cite{suppl}). We are now  going to consider the wavepacket motion in these two situations. 

\emph{Crossing bands: Rashba SOC (Dirac cone).} 
The equation of motion~\eqref{longeq2} writes explicitly:
\begin{eqnarray}
\Dot{\bm{r}}  &=& \left( {\frac{{{\hbar }{k}}}{{m}} + \frac{\alpha}{\hbar} \cos \left( {{\phi _s} - {\phi _k}} \right)\sin  {{\theta _s}} } \right)\left( \begin{gathered}
  \cos \,{\phi _k} \hfill \\
  \sin \,{\phi _k} \hfill \\ 
\end{gathered}  \right)\nonumber\\ &+& \Omega \sin\theta_s\sin \left( {{\phi _k} - {\phi _s}} \right)\left( \begin{gathered}
  \sqrt {{g_{{k_x}{k_x}}}}  \hfill \\
  \sqrt {{g_{{k_y}{k_y}}}}  \hfill \\ 
\end{gathered}  \right)
\label{Rashbasol}
\end{eqnarray}
where $\phi_k$ is the polar angle of the wavevector, to which the effective field is antiparallel ($\phi_f=\phi_k-\pi$). This equation contains only the orientation of the spinor $\theta_s,\phi_s$ and the center of mass wavevector $k$. The first part of the group velocity contains the spin-independent parabolic dispersion and a spin-dependent contribution, with the propagation direction ultimately controlled by the current orientation of the spinor. The second part of this expression, which includes the QM $g_{\bm{kk}}$, appears because of the explicit time dependence of the spinor. The $x$ and $y$ projections of the velocity are controlled by the corresponding projections of the QM.

As an illustration, we consider the case without external fields (${\Dot{k}}=0$), with $\bm{k}=k_0\bm{e}_x$ ($\phi_k=0$), and the spinor $\bm{S}=S_z\bm{e}_z$ perpendicular to the effective field at $t=0$, so $\theta_s=0$. The wavefunction is projected equally on both bands, and the pseudospin precession frequency is $\Omega=2\alpha k_0/\hbar$.  

In this case, the $x$ projection of the group velocity is constant. The time-dependent trajectory reads: 
\begin{eqnarray}
x(t) &=&  \frac{\hbar k_0}{m}t \\
y(t) &=& (1-\cos\Omega t)\sqrt {{g_{{k_y}{k_y}}}}=\frac{1-\cos\Omega t}{2 k_0}\nonumber
\end{eqnarray}
because the QM is ${g_{{k_y}{k_y}}} = 1/4 k_0^2$. This oscillating motion due to the pseudospin precession is the ZBW effect. The magnitude of the oscillation along $y$ is given by $\sqrt {g_{{k_y}{k_y}}}$, which acts as a fundamental characteristic length scale of the problem, as we will discuss more in details below.

To confirm our analytical results, we perform numerical simulations, solving the 2D spinor Schr\"odinger equation
\begin{equation}
    i\hbar\frac{\partial \psi}{\partial t}=\hat{H}\psi
    \label{Schro}
\end{equation}
with the Rashba Hamiltonian \eqref{RashbaHam}, taking a finite-size wavepacket centered at a wavevector $k_0$ with its spinor part given by $(1,0)^T$ (corresponding to $\theta_s=0$) as an initial condition. We choose the simulation parameters typical for polaritonic systems \cite{gianfrate2020measurement}: $\alpha=1$~meV/$\mu$m$^{-1}$, $k_0=1$~$\mu$m$^{-1}$, $m=2\times 10^{-5}m_0$ ($m_0$ is the free electron mass). We can observe a truly excellent agreement with the analytical trajectory in a substantial time window, limited only by the transient nature of the ZBW due to the finite wavepacket size~\cite{Lock1979}.

As said in the introduction, the Rashba Hamiltonian can be described as resulting from the action of a non-Abelian gauge field \cite{Shen2005,Jin2006,Hatano2007,Yang2008} described by the Yang-Mills Lagrangian~\cite{Yang1954}. Within this picture,
it is also possible to derive a semiclassical equation of motion, where the acceleration is the result of the action of a non-Abelian magnetic force acting on (pseudo)-spin currents, as recently measured in \cite{fieramosca2019chromodynamics}. As shown in the Supplementary \cite{suppl}, the time derivative of Eq.\eqref{Rashbasol} gives an expression of the transverse acceleration in terms of the QM, equivalent to the results of the Yang-Mills theory \cite{Jin2006}. This acceleration appears because of the precession of the spin, or, in other words, because of the interband transitions described by the QM. This provides a microscopic mechanism behind the non-Abelian Lorentz force of the Yang-Mills field, which can be interpreted as being the consequence of the geometry of the underlying quantum space.

\begin{figure}[tbp]
\begin{center}
\includegraphics[width=1\linewidth]{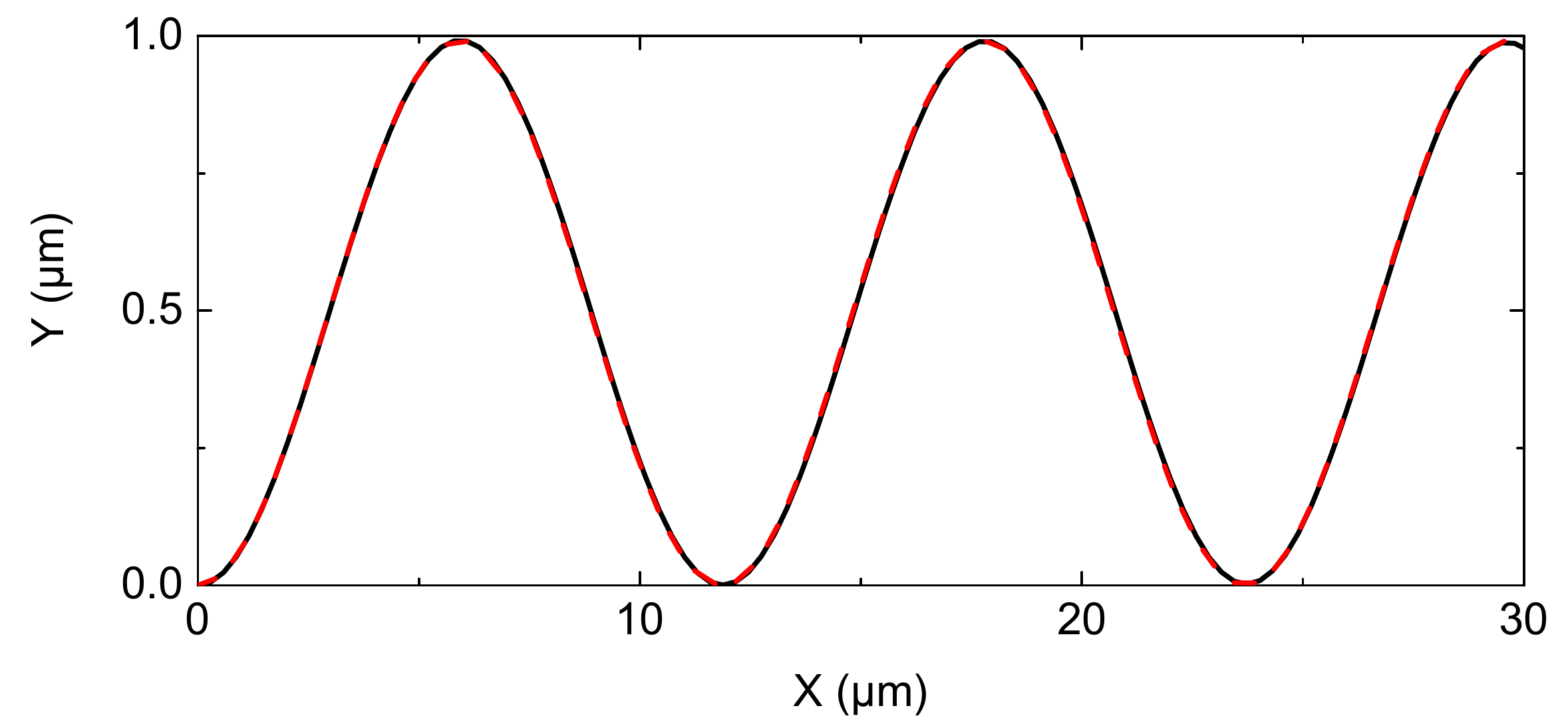}
\caption{Rashba Hamiltonian. Wavepacket  dynamics  from  Schr\"odinger equation  (black solid lines) and analytical solution of semiclassical  Eqns.~\eqref{dsdqdt},\eqref{longeq} (red dashed lines). \label{fig2}}
\end{center}
\end{figure}

\begin{figure}
    \centering
    \includegraphics[width=0.99\linewidth]{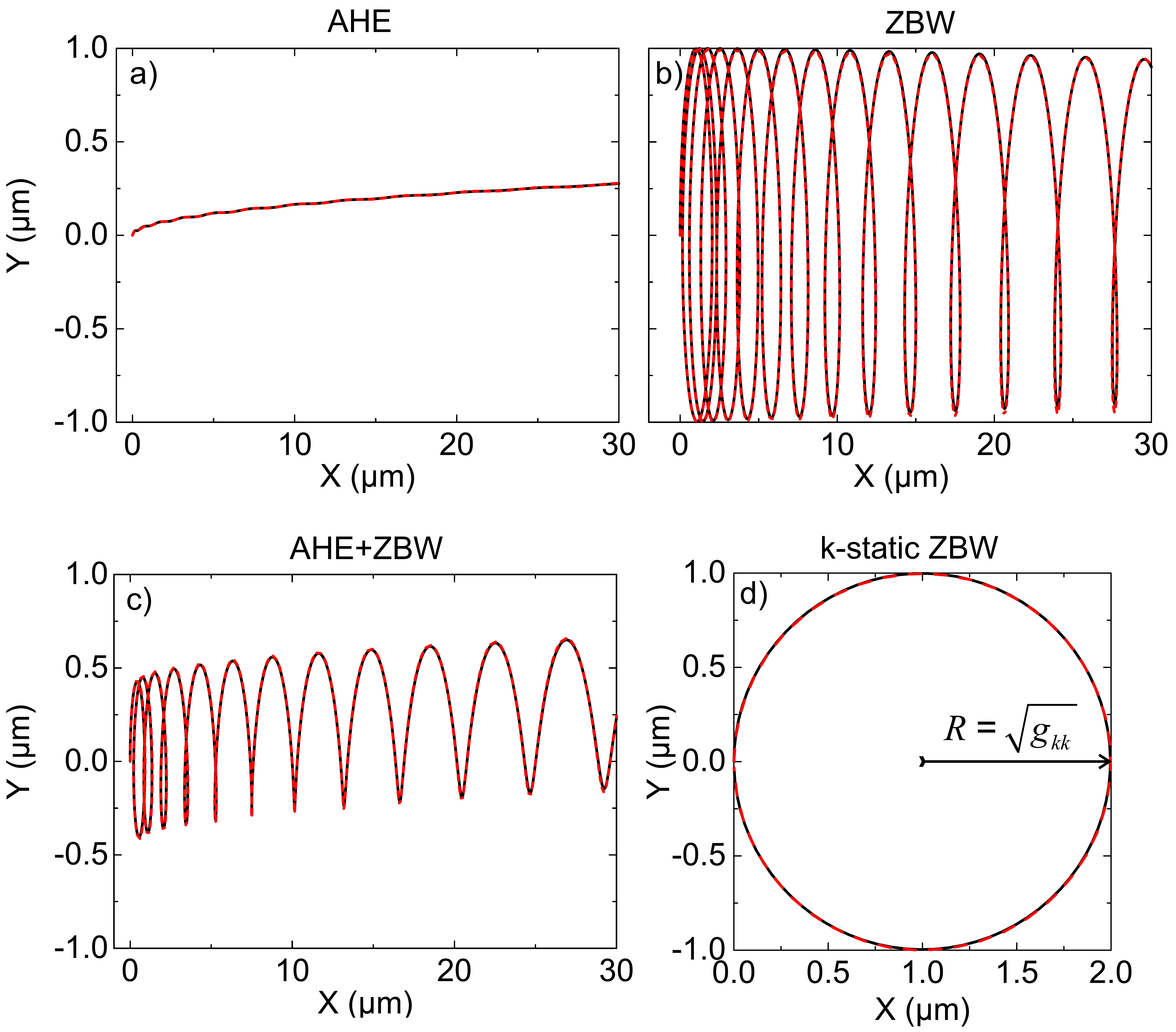}
    \caption{Rashba+Zeeman Hamiltonian. Wavepacket dynamics from Schr\"odinger (black solid lines) and semiclassical (red dashed lines) equations: a) AHE (single-band initial excitation); b) ZBW (equal fractions of both bands); c) both effects together; d) cyclotron-like orbits at constant $k$ (no potential gradient) with a radius determined by the metric.}
    \label{fig3}
\end{figure}

\emph{Anticrossing bands: Rashba + Zeeman (massive Dirac).} 
We now consider the Rashba Hamiltonian combined with a Zeeman term. The resulting bands are non-degenerate and show a distributed Berry curvature. A wavepacket accelerated in such a system can show either AHE or ZBW, or a combination of both effects, depending on initial conditions. 
Figure~\eqref{fig3}(a-c) considers the acceleration by a spatial energy gradient $2\times 10^{-3}$~meV/$\mu$m for different initial conditions. We compare the center of mass trajectories obtained by solving the spinor Schr\"odinger equation \eqref{Schro} and the one obtained from the semiclassical equations of motion \eqref{longeq}. The Zeeman splitting is $\Delta=0.5$~meV, other parameters as above. Panel (a) demonstrates the AHE regime, with the initial condition corresponding to an eigenstate of the system (the lowest energy band at $k=0$): the deviation along $y$ is the AHE drift. The correspondence between the description of the AHE in terms of Berry curvature and the one based on the use of the QM is explicitly shown in the supplemental material \cite{suppl}. Panel (b) corresponds to the pure ZBW, with the initial condition corresponding to the equal fraction of both branches: $f_1=f_2$. In this case, there is no AHE drift, because the effect of the Berry curvature is completely canceled by $f_1-f_2=0$. Finally, panel (c) corresponds to a particular case of $f_1-f_2=0.9$, allowing to observe both the AHE drift and the large oscillations due to the ZBW. Qualitatively similar results are obtained with a TE-TM SOC \cite{suppl} characterized by a double winding number and typical for photonic systems.
The AHE has been recently measured in an optical system  with TE-TM SOC and Zeeman splitting \cite{gianfrate2020measurement}. 

If we consider a wavepacket with zero intial wavevector and zero external force $\dot{k}=0$, the effective field is completely determined by the Zeeman splitting: $\theta_f=0$. If the spin of the initial wavepacket is in the plane $\theta_s=\pi/2$ (and $\phi_s=0$ as an example), the third equation of motion~\eqref{dsdqdt} gives that the spin will remain in the plane ($\sin\theta_s=1$) and rotate  with an angular frequency $\Omega$ ($\phi_s=\Omega t$). The equation~\eqref{longeq} gives:
\begin{equation}
\Dot{x}=\Omega\sqrt{g_{k_x k_x}}\cos\Omega t,\quad 
\Dot{y}=\Omega\sqrt{g_{k_y k_y}}\sin\Omega t
\end{equation}
where the QM at $k=0$ is given by $g_{k_x k_x}=g_{k_y k_y}=\alpha^2/4\Delta^2$
These equations explicitly show the wavepacket rotation in real space, with a  radius  determined by the value of the QM $R=\sqrt{g_{\bm{kk}}}=\alpha/2\Delta$, as illustrated in Fig.~\ref{fig3}(d). 

Our results show that the QM provides a characteristic length scale $l=\sqrt{g_{max}}$ for the semiclassical behavior. This is best seen with the example of the Dirac equation, where the value of the QM at $k=0$ is
\begin{equation}
    \sqrt{g_{kk}}=\frac{\hbar}{mc},
\end{equation}
which is the well-known Compton wavelength $\lambda_C$ of the electron, determining a universal length scale in Physics. Indeed, it enters the expressions for the classical electron radius, the Bohr radius, the electron-proton scattering cross-section, and even determines the Planck length. 

The physical meaning of the Compton wavelength can be understood with the QM. Qualitatively, it limits the precision of the measurement of the electron's position. Indeed, scattering of a photon with the wavelength $\lambda_C$ brings the electron into a 50\% electron-positron superposition, corresponding exactly to the case of Fig.~\ref{fig3}(d): the electron's center of mass exhibits a  cyclotron motion with the radius $R=\sqrt{g_{kk}(0)}=\lambda_C$. This rotation is what determines the uncertainty of its position.  Even for Hamiltonians which do not have a single length scale, the QM still can be used to determine the scales of the ZBW at rest or at high velocities, changing from the Compton to the de Broglie wavelength \cite{Bolte2007,Zawadzki2011}: indeed, $\sqrt{g_{kk}}\sim\ 1/k=\hbar/p$ at large $k$.

The value of the QM at its extremum has a very fundamental nature. Since the integral of the QM is often quantized, representing a topological invariant \cite{mera2021k} similar to the Chern number, this maximal value also determines the extension of the metric in the parameter space, that is, the characteristic scale at which the changes occur (for example, level crossing). It determines both the maximal amplitude of the ZBW oscillations and of the anomalous Hall drift (even though the latter is an integral quantity). It determines the spatial extension of the chiral edge state in topological insulators, controlling the minimal size of topological lasers and optical isolators. This will be a subject for future works.

\emph{Conclusions} We derived the semiclassical equations of motion for a wavepacket in a multiband system in terms of the static band parameters, in particular, the QM. The latter turns out to determine a universal length scale for all effects beyond the simple group velocity.

\begin{acknowledgments}
We acknowledge useful discussions with V. Rabant. We acknowledge the support of the projects EU "QUANTOPOL" (846353), "Quantum Fluids of Light"  (ANR-16-CE30-0021), of the ANR Labex GaNEXT (ANR-11-LABX-0014), and of the ANR program "Investissements d'Avenir" through the IDEX-ISITE initiative 16-IDEX-0001 (CAP 20-25). 
\end{acknowledgments}

\bibliography{biblio}

\addto\captionsenglish{\renewcommand{\figurename}{Supplementary Figure}}
\addto\captionsenglish{\renewcommand{\refname}{Supplementary References}}
\renewcommand{\theequation}{S\arabic{equation}}
\setcounter{equation}{0}

\section{Supplemental Material}
In this Supplemental material, we provide more details for the calculation of the spherical angles, discuss the contribution of the quantum metric and the Berry curvature to the semi-classical equations, compare our set of equations to the previous multi-band formalism, and estimate the maximal values of the ZBW and AHE effects.

\subsection{Spherical angles and the quantum metric}

The equation for the angle $\eta$ given by the spherical cosine law reads:
\begin{equation}
    \cos\eta=\cos\theta_s\cos\theta_f+\sin\theta_s\sin\theta_f\cos\left(\phi_f-\phi_s\right)
    \label{sphercos}
\end{equation}
The contribution to the spin-dependent part of the energy stemming from the wavevector dependence of the coefficients $f_i$ can be found via the wavevector dependence of the spherical coordinates $\theta_f,\phi_f$ of the effective field:
\begin{equation}
    \frac{\partial f_i}{\partial k_j}=\frac{\partial f_i}{\partial \phi_f}\frac{\partial \phi_f}{\partial k_j}+\frac{\partial f_i}{\partial \theta_f}\frac{\partial \theta_f}{\partial k_j}
\end{equation}
and the latter are determined by the metric part of the  quantum geometric tensor (QGT) \cite{provost1980riemannian}.
 This allows writing
\begin{equation}
    \frac{\partial \theta_f}{\partial k_i}=2\sqrt{g_{k_i k_i}}\cos\zeta,\quad \frac{\partial \phi_f}{\partial k_i}=2\sqrt{g_{k_i k_i}}\frac{\sin\zeta}{\sin\theta_f}
\end{equation}
with $\zeta$ controlled by the evolution of the effective field with $k_i$ as $\tan\zeta=\sin\theta_f(\partial\phi_f/\partial\theta_f)_i$. 

We note that a non-zero Berry curvature requires the effective field to cover a solid angle on the Bloch sphere, that is, the effective field should move along different axes, which means different values of $\zeta$ for $k_x$ and $k_y$. Different terms will therefore appear in the equations for the two projections of the velocity.

\subsection{Berry curvature and quantum metric}

The one-to-one correspondence between the Berry curvature and the QM was discussed in Ref.~\cite{Piechon2016}. Given the existence of such mapping, it is therefore natural that it is possible to write the semi-classical equations using either the Berry curvature or the QM. Here, we demonstrate that the Berry curvature term responsible for the AHE drift gives exactly the same contribution to the transverse group velocity as the term written in the equation using the QM.

In the particular case of the Rashba SOC with the Zeeman splitting we consider as an example, the AHE drift occurs in the $y$ direction. Therefore, we need to study the $y$ projection of the group velocity. To establish the equivalence between the equations with the Berry curvature, containing the time derivative of the wavevector $dk_y/dt$ and the semi-classical equations with the QM, we will use the description of the non-adiabaticity  by the QM. 

We begin by providing the explicit expressions for the Berry curvature and the quantum metric for the Rashba/Zeeman Hamiltonian, equivalent to the massive Dirac Hamiltonian. The Berry curvature reads:
\begin{equation}
    B_z=\frac{\alpha^2\Delta}{2\left(\Delta^2+\alpha^2 k^2\right)^{3/2}}
\end{equation}
and the QM reads
\begin{eqnarray}
g_{k_x,k_x}&=&\frac{\alpha^2\left(\Delta^2+\alpha^2 k_y^2\right)}{4\left(\Delta^2+\alpha^2 k^2\right)^2}\\
g_{k_y,k_y}&=&\frac{\alpha^2\left(\Delta^2+\alpha^2 k_x^2\right)}{4\left(\Delta^2+\alpha^2 k^2\right)^2}
\end{eqnarray}
The term of the equation (6) of the main text responsible for the transverse anomalous Hall velocity (for a wavepacket characterized by a wavevector along $x$) reads:
\begin{equation}
    \hbar v_{y}=\ldots
    +\hbar\Omega\sqrt{g_{k_y,k_y}}\sin\theta_s\sin(\phi_f-\phi_s)
\end{equation}
In the quasi-adiabatic regime,  $\sin\theta_s\approx\sin\theta_f=\alpha k/\hbar\Omega$ and $\sin(\phi_f-\phi_s)\approx \eta/\sin\theta_f$, which gives:
\begin{equation}
    \hbar v_{y}=\ldots
    +\hbar\Omega\sqrt{g_{k_y,k_y}}\eta 
\end{equation}
The angle $\eta$ here can be obtained from the wavevector change rate $dk_x/dt$ using the fact that any non-zero change of the parameters of the Hamiltonian leads to a finite non-adiabaticity described by $\eta$ and given by the quantum metric along the wave vector evolution direction \cite{Bleu2018}:
\begin{equation}
    f_{NA}=\frac{g_{k_x,k_x}}{\hbar^2\Omega^2}\left(\frac{dk_x}{dt}\right)^2
\end{equation}
This non-adiabatic fraction is linked with the angle between the spin and the effective field $\eta$ as $f_{NA}=\eta^2/4$. 
This allows us to transform the expression obtained using the QM to the familiar expression with the Berry curvature:
\begin{equation}
\hbar v_y=\ldots+2\sqrt{g_{yy}g_{xx}}\frac{dk_x}{dt}=B_z\frac{dk_x}{dt}
\end{equation}
where we have used the identity
\begin{equation}
    \sqrt{\det g}=\frac{B_z}{2}
\end{equation}
valid for all 2-band Hamiltonians with a Berry cuvature of a constant sign \cite{mera2021k}. The two approaches indeed give the same contribution to the transverse velocity.

\subsection{Comparison with Culcer's equations}
The semi-classical equations of motion for a wave-packet  with the center of mass position $\bm{r}_c$ and wavevector $\bm{q}_c$, were written in Ref.~\cite{Culcer2005}:
\begin{eqnarray}
\hbar\mathbf{\Dot{q}}_c&=&-\frac{\partial E}{\partial \mathbf{r}_c}-\mathbf{B}_{t\mathbf{r}}\label{ksp}\\
\hbar\mathbf{\Dot{r}}_c&=&\frac{\partial E}{\partial \mathbf{k}_c}-\mathbf{B}_{\mathbf{qq}}\mathbf{\Dot{q}_c}+\mathbf{B}_{t\mathbf{q}}\label{rsp}\\
i\hbar\frac{dc_i}{dt}&=&\left(H_{ij}-\hbar\bra{u_i}\ket{i\frac{du_j}{dt}}\right)c_j\label{pseudo0}
\end{eqnarray}
with the coefficients $c_i$ determining the composition of the wavepacket within the bands with the numbers $i$, $u_i$ being periodic spinor part of the wave-packet wave function,  $\mathbf{B}_{\mathbf{qq}}$ is the Berry curvature tensor, calculated not for a single band, but for a superposition of bands, and
\begin{equation}
    {B}_{t\mathbf{q}}^\alpha=i\left(\bra{\frac{\partial u}{\partial t}}\ket{\frac{\partial u}{\partial q_\alpha}}-\bra{\frac{\partial u}{\partial q_\alpha}}\ket{\frac{\partial u}{\partial t}}    \right)
\end{equation}
is an additional Berry curvature tensor  $\mathbf{B}_{t\mathbf{q}}$ which appears because of the explicit time dependence $u(t)$ (absent in the case of a single band).

\subsection{Quantum metric and the non-Abelian gauge field}

A general non-relativistic Hamiltonian of a massive matter field (quantum particle) minimally coupled with a non-Abelian gauge field determined by a vector potential $A_\mu^a$ reads \cite{Yang1954,Ryder,Jin2006}:
\begin{equation}
    H_{YM}=\frac{1}{2m}\left(\hat{\bm{p}}-\eta\bm{A}^a\sigma^a\right)^2+\eta A_t^a\sigma^a
    \label{HYM}
\end{equation}
The coupling constant is $\eta=\hbar/2$ (the quantum of spin). We use upper number indices $0-3$ for Pauli matrices. Comparing this expression with the Rashba Hamiltonian [Eq. (8) of the main text], one sees that only two components of the vector potential are non-zero: $A_x^1=-m\alpha/\eta$, $A_y^2=-m\alpha/\eta$. 
The non-Abelian nature of the field makes that the constant vector potential nevertheless gives rise to non-zero field strength tensor. In the case of Rashba SOC, the only non-zero components are $F_{yx}^3=-F_{xy}^3=-m^2\alpha^2/g$, where $g=\hbar/2$ is the Yang-Mills coupling constant (upper indices $0-3$ correspond to Pauli matrices). This non-zero field is responsible for an analogue of a Lorentz force for a non-Abelian gauge field. The Yang-Mills theory thus allows to predict an analogue of a transverse force acting on a spin current in the Rashba Hamiltonian. This force is proportional to the field strength and to the spin current, as can be seen from the second  Newton's law:
\begin{equation}
 m\,dv^\mu/d\tau=\bm{J}_\nu\cdot\bm{F}^{\mu\nu}
 \label{chrome}
\end{equation}
The corresponding acceleration is ultimately found as $a_x=-2m\alpha^2J_y^3/\hbar^2$, $a_y=2m\alpha^2J_x^3/\hbar^2$, where $J_x^3$, $J_y^3$ are the circular (spin-up/down) components of the spin current propagating along $x$ and $y$ respectively.

We will now compare the predictions of the Yang-Mills theory with those of the semi-classical equations that we have derived. In the particular case where the external forces are absent, the first of the equations of motion (5) of the main text gives that the central wavevector of the wave-packet is constant: $\Dot{\mathbf{k}}=0$. The equation (6) of the main text is still time-dependent, so it can be derived once again to find an analogue of the second Newton's law, similar to \eqref{chrome}.  We consider a parabolic band extremum characterized by an effective mass $m$, and define the $z$-projection of the spin current as $\mathbf{J}=\hbar^2 \mathbf{q}_c\cos\theta_s/2m$,  which allows writing
\begin{equation}
     m\Ddot{\mathbf{r}}=\sqrt{g_{\mathbf{kk}}}\frac{4\alpha^2 k m}{\hbar^2}\mathbf{e_z}\times \mathbf{J}
     \label{YMF}
\end{equation}
making the metric appear explicitly in the expression for the non-Abelian magnetic-like Yang-Mills force. We can therefore conclude that the QM is at the heart of the microscopic mechanism behind the Lorentz-like transverse force acting on a spin current in the static non-Abelian gauge field described by Eq.~\eqref{chrome}.

The covariant derivative appears in the Lagrangian to ensure the fundamental principle of gauge invariance. But the  physical mechanism associated with its microscopic effect is based on the fact that the group velocity in a spinor system necessarily includes the QM describing the interband transitions due to the spin dynamics.

\subsection{Estimation of the maximal values of the effects}
We note that the scale of both the AHE and the ZBW is quite comparable in both configuration (TE-TM or Rashba SOCs + Zeeman splitting), both in what concerns the maximal lateral deviation, maximal velocity, or maximal acceleration. In general, the ZBW is larger above a certain wavevector. We note that the combination of the Rashba SOC with the Zeeman splitting is equivalent (in what concerns its spinor part) to the gapped Dirac Hamiltonian. This makes the applicability of our example even broader. 

The pure AHE determined by the Berry curvature requires adiabatic evolution of the system. The additional group velocity due to AHE is given by
\begin{equation}
    \Dot{\bm{q}}_{\mathrm{AHE}}=\bm{B}\times\Dot{\bm{k}}
\end{equation}
and if the potential gradient is constant, the acceleration due to AHE can be found as
\begin{equation}
    \Ddot{\bm{q}}_{\mathrm{AHE}}=\Dot{\bm{B}}\times\Dot{\bm{k}}
\end{equation}
We consider a wave-packet accelerated from $k=0$ along the $x$ axis. The transverse acceleration therefore can be found as
\begin{equation}
    \frac{d^2r_y}{dt^2}=\frac{dB_z}{dk_x}\left(\frac{dk_x}{dt}\right)^2
\end{equation}
To find the maximal acceleration we need to find the maximal value of $dB_z/dk_x$, which occurs at $k_{max}\approx 0.48\sqrt{\Delta/\beta}$, which gives 
\begin{equation}
    \left(\frac{dB_z}{dk_x}\right)_{max}\approx 1.51 \frac{\beta^{3/2}}{\Delta^{3/2}}
\end{equation}
On the other hand, the maximal value of the acceleration $dk_x/dt$ acceptable for the adiabatic evolution can be obtained from the expression for the non-adiabatic fraction \cite{Bleu2018}:
\begin{equation}
    f_{\mathrm{NA}}=\frac{g_{kk}}{\Omega^2}\left(\frac{dk_x}{dt}\right)^2
\end{equation}
Requiring that the non-adiabatic fraction does not exceed 20\% gives the maximal value of the square of the derivative
\begin{equation}
    \left(\frac{dk_x}{dt}\right)^2_{max}\approx \frac{\Delta^3}{\hbar^2 \beta}
\end{equation}
Qualitatively, the same condition can be obtained from the physical requirement that the energy of the eigenstate should not change faster than the value of the frequency corresponding to the splitting between the eigenstates.

Putting both expressions together, we find
\begin{equation}
    a_{\mathrm{AHE,max}}\approx 1.5 \frac{\beta^{1/2}\Delta^{3/2}}{\hbar^2}
\end{equation}

On the other hand, the maximal acceleration due to the QME depends on the wavevector, which for the system with TE-TM SOC (and without the Zeeman splitting) gives:
\begin{equation}
    a_{\mathrm{QME,max}}=\Omega^2\sqrt{g_{yy}}=\frac{4\beta^2 k^3}{\hbar^2}
\end{equation}

Comparing both expressions, we see that for $k>2^{-2/3}\sqrt{\Delta/\beta}$, the QME acceleration is larger than the AHE one, whereas for the wavevector where the AHE acceleration is maximal, it is approximately twice larger than the QME: $a_{\mathrm{AHE},max}\approx 2a_{\mathrm{QME}}(k_{max})$. The QME acceleration can be increased unlimitedly by increasing the wavevector $k_x$.

For the configuration with Rasbha SOC and Zeeman splitting, the accelerations are given by $a_{\mathrm{AHE},max}\approx 0.43\alpha\Delta/\hbar^2$ and $a_{\mathrm{QME,max}}=2\alpha^2 k/\hbar^2$. For the same wavevector $k_{\mathrm{max}}=\Delta/2\alpha$ (giving the maximal AHE acceleration), one finds $a_{\mathrm{AHE,max}}\approx 0.43 a_{\mathrm{QME,k_max}}$. The AHE acceleration is therefore smaller than the QME one at this wavevector. However, for this configuration, it is better to compare the transverse velocities, and not the accelerations, because the AHE acceleration is actually reducing the effect (there is a non-zero transverse velocity at $t=0$, which then decays). 

The expressions for the maximal velocities give the same result:
\begin{equation}
    v_{\mathrm{AHE,max}}=v_{\mathrm{QME,max}}=\frac{\alpha}{2\hbar}
\end{equation}

It is also interesting to compare the maximal lateral shift in both cases. Both AHE and QME shifts exhibit a divergent behavior on a certain parameter ($\Delta_Z$ for AHE, $k_0$ for QME), and can, in principle, exhibit arbitrarily large values, if these two parameters are decreased. On the other hand, the experimental observations are limited by the line broadening $\Gamma$ (due to finite lifetime) and by the wave-packet size in real space. For the Rashba Hamiltonian, the broadening sets a common limit for both effects: $\Delta Y_{\mathrm{AHE/QME,max}}\sim \alpha/\Gamma$. The wave-packet size also sets a similar limit: $\Delta Y_{\mathrm{AHE/QME,max}}\sim 1/\sigma_k\sim \sigma_r$: the maximal scale of both effects is determined by the wave-packet size. For AHE, this condition means the WP should be smaller than the region with non-zero Berry curvature, and for QME that $k_0$ is really different from 0.

We conclude that while the two effects occur in completely different regimes (adiabatic vs coherent oscillations), their scales are actually quite comparable.

\end{document}